# Nonlinear and thermo-optical characterization of bare *imidazolium* ionic liquids

**Vinícius C. Ferreira[1], Graciane Marin[2], Jairton Dupont[2] and Ricardo R. B. Correia[1]**

[1] OPTMA - Optics, Photonics and Materials Group, Institute of Physics, Universidade Federal do Rio Grande do Sul, Porto Alegre, Brazil
[2] LAMOCA - Laboratory of Molecular Catalysis, Institute of Chemistry, Universidade Federal do Rio Grande do Sul, Porto Alegre, Brazil

E-mail: vinicius.ferreira@ufrgs.br, rego@if.ufrgs.br

**Abstract**

Nonlinear optical and thermo-optical properties of two pure ionic liquids, $BMIOMe.NTf_2$ and $BMIOMe.N(CN)_2$, were examined here. It was the first nonlinear refractive index determination of a pristine ionic liquid by a standard self-refraction experiment. The nonlinear optical characterizations were performed using Z-scan and EZ-scan techniques in the thermally managed approach with a mode-locked femtosecond laser source. Thermal properties were analyzed concomitantly and the thermo-optical coefficient, thermal characteristic time, and lens strength were characterized. These results show that $BMIOMe.NTf_2$ is a prominent material to be engineered for photonics applications.

## 1. Introduction

Low temperature molten salts known as room temperature ionic liquids (RTILs) have been emerged as green solvents regarding to the environmental concerns. Among them, imidazolium-based ILs are very interesting fluids whose hydrophobic and hydrophilic nature can easily be tuned by simply exchanging their counteranoins [1]. These ILs have gain much attention because most of them display negligible vapour pressure, high chemical and thermal stability, a wide electrochemical window, non-flammability, melting points below 100 °C, high ionic conductivity, high boiling points and stability [2]. Because of their negligible vaopur pressure, these fluids are using as a working media under ultra vacuum conditions (10-9 mbar), and can expose them to an electron and/or ion beam without the charging effect [3,4]. Si (100) anchored imidazolium ILs derived microstructures that represented light transmission properties and acted as photonic elements for near-infrared [5,6].

Considering the photonic applications, optical and photothermal characterizations are essential, as well as the associated nonlinear optical properties. Nonlinear optics is a regime where high orders of susceptibility are excited, and specif phenomena are generated, in which the optical properties change with the presence of light. These high orders are achieved typically with pulsed lasers source because they have a high peak intensity. Those effects are a response of an asymmetric deformation in the charge distribution or the polarizability of a molecule [7]. Nonlinear optical, NLO, properties have been a way to achieve distinct possibilities in the technologic evolution, which show itself essentially in several areas, as signal processing, medical, and laser science [8]. In this way, ILs have been studied and showed as reasonable material with NLO response since they usually present an asymmetry charge distribution [9].

There are several techniques for NLO characterization. In this work we employ the technique of scanning the sample in the direction of the beam propagation, typically Z-scan and EZ-scan. These techniques can measure NLO refractive and absorptive properties, and thermo-optical properties of materials [1,2]. We can split the effects measured by these techniques into two main groups: one is the nonlinear refractive/electronic response, also called as local or fast response, which is generated by the nonlinear refractive index excitation; and the thermal response, sometimes called as nonlocal or slow response, it came from the sample heating. Several studies were performed in IL yet. Nonlinear refractive indexes by the Z-scan technique were studied and characterized for BMI.BF4 with colloidal solution of silver nanoparticles dispersed in [3] , mixtures of azobenzene containing ILs crystalline polymer [4] and thin film of IL crystalline polymer [5]. The range of thermal response is bigger since this response is typically much stronger than the refractive. In this scope, BMI.BF4 and BMI.PF6 were studied in a laser source of continuous-wave and femtosecond pulses [6], several ILs also with a high repetition femtosecond source [7], IL doped with nanoparticles in a nanosecond pulse laser source [8], and twelve different ILs in a setup illuminated by a continuous-wave Ar laser [9].

In this paper we present a study of two ILs 1-(2-methoxy-ethyl)-3-methyl imidazolium (trifluoromethylsulfonyl) imidate, BMIOMe.NTf$_2$, and 1-(2-methoxy-ethyl)-3-methyl imidazolium dicyanamide, BMIOMe.N(CN)$_2$. These ILs were characterized by Z-scan and EZ-scan technique, in a setup using a femtosecond laser source, and the nonlinear refractive indexes and thermo-optical properties were characterized. We introduce the first value obtained experimentaly for the NLO refractive index of a pure IL by a standard self-refraction experiment. Concurring thermal lens effects were observed and related thermo-optical parameters were also evaluated based on linear absorption spectra measured for these samples.

## 2. Ionic liquids

Two ILs were studied and characterized in this work, BMIOMe.NTf$_2$, 1-(2-methoxy-ethyl)-3-methyl imidazolium (trifluoromethylsulfonyl)imidate, and BMIOMe.N(CN)$_2$, 1-(2-methoxy-ethyl)-3-methyl imidazolium dicyanamide. The molecular structures of these ILs are presented in figure 1.

Solvents used in the synthesis were purchased from commercial sources and were used as received. 1-Methylimidazole was distilled under reduced pressure. All other reagents were used without further purification. It was performed nuclear magnetic resonance (NMR) spectroscopy. 1H and 13C NMR spectra were obtained on Varian Inova 400 and 500 MHz spectrometers using (CD3)2CO as the solvent, presented in figure 2 and 3. The purity of these liquids can be determined by NMR spectra [10].

*2.1 Synthesis of* BMIOMe.NTf$_2$. A solution of 1-methoxyethane-3-methylimidazolium methanesulfonate (50 g in 40 mL of water) is added to a solution of lithium bis (trifluoromethanesulfonyl) imidate (LiN (Tf2)) (19.33 g in 10 mL of water) and the reaction occurs on the order of 1: 1 with 5% excess LiN (Tf2). The mixture is kept under stirring for two hours and then washed with 100 mL of dichloromethane to remove water. The dichloromethane is removed with the aid of a rotary evaporator and finally dried under vacuum. 1H NMR (500 MHz, acetone) δ 8.94 (s, 1H), 7.69 (s, 1H), 7.65 (s, 1H), 4.50 (t, J = 5.0 Hz, 2H), 4.05 (s, 3H), 3.80 (t, J = 5.0 Hz, 2H), 3.34 (s, 3H). 13C NMR (126 MHz, acetone) δ 137.66, 120,85 (q, JC-F = 320Hz, CF3), 124.41, 123.85, 70.71, 58.79, 50.36, 36.60.

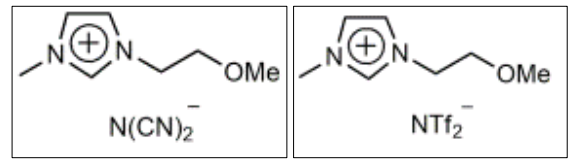

Figure 1: Molecular structure of ionic liquids studied here, BMIOMe.NTf$_2$ and BMIOMe.N(CN)$_2$.

*2.2 Synthesis of* BMIOMe.N(CN)$_2$. A solution of 1-methoxyethane-3-methylimidazolium methanesulfonate (10 g in 10 mL of methanol) was added to a solution of sodium dicyanamide (NaN (CN)2) (4.9 g in 107 mL). Previously, the dicyanamide solution was heated until the complete dissolution of the solute. Stirring was continued for 15 minutes. The vessel was brought to an ice bath for 15 minutes to precipitate NaCH3SO3 which was filtered off. A further 1 g of NaN(CN)2 was added, the solution was again stirred for 15 minutes, then the cooling and filtration process was repeated. Subsequently, the methanol was evaporated and the ionic liquid extracted with dichloromethane. Celite and active carbon were used in the final purification. Yield 77.6%. Degree of purity: 99%. 1H NMR (400 MHz, acetone) δ 9.10 (s, 1H), 7.74 (t, J = 1.7 Hz, 1H), 7.70 (t, J = 1.7 Hz, 1H), 4.51 (t, J = 5.0 Hz, 2H), 4.05 (s, 3H), 3.80 (t, J = 5.0 Hz, 2H), 3.34 (s, 3H). 13C NMR (100 MHz, acetone) δ 137.85, 124.31, 123.75, 120.42, 70.65, 58.77, 50.16, 36.50.

## 3. Nonlinear optics setup

Nonlinear refractive index and thermal optical analyses were evaluated by the EZ-scan and Z-scan technique respectively. The fundamental difference between these techniques is the shape of the spatial filter used to select regions of the beam transmitted by the sample. These methods rely on the NLO self-focalization effect induced by a laser beam within the sample. When a thin sample is scanned through the Rayleigh range of a focused beam, the associated wavefront undergoes an auto induced curvature change that depends on the relative position of the sample from the focal plane. A detector is placed allowing to collect all the transmitted beam power and, depending on the selected technique, appropriate spatial filters are positioned in front of the detector. Wavefront changes induced by the laser beam are proportional to the nonlinear and thermo-optical properties of the sample. The variable wavefront curvature alters the transmittance through the spatial filter and, therefore, also the amount of detected light. In the case of Z-scan, a hole is used as the spatial selector and thermal and electronic effects properties of the materials can be evaluated [1,11]. Using a disc instead of a hole as the spatial filter, the technique called EZ-scan, the sensitivity and the signal noise increase, however, there is no well-established or a simple model to define the thermo-optical properties [12–14]. A third variation of the experimental setup can be used to evaluate the nonlinear absorption. The technique called Open Aperture Z-scan is performed by removing the spatial filter in front of the detector. In this case, all variation of transmittance will occur just in the case of light appear or disappear during the performance of scan [1]. The laser source in our setup is a mode-locked Ti:Sapphire laser oscillator ($76\ MHz$, $150\ fs$, $780\ nm$) Mira 900 by Coherent. The experimental arrangement was previously presented alongside a procedure to reduce noise and systematic error in the data [15].

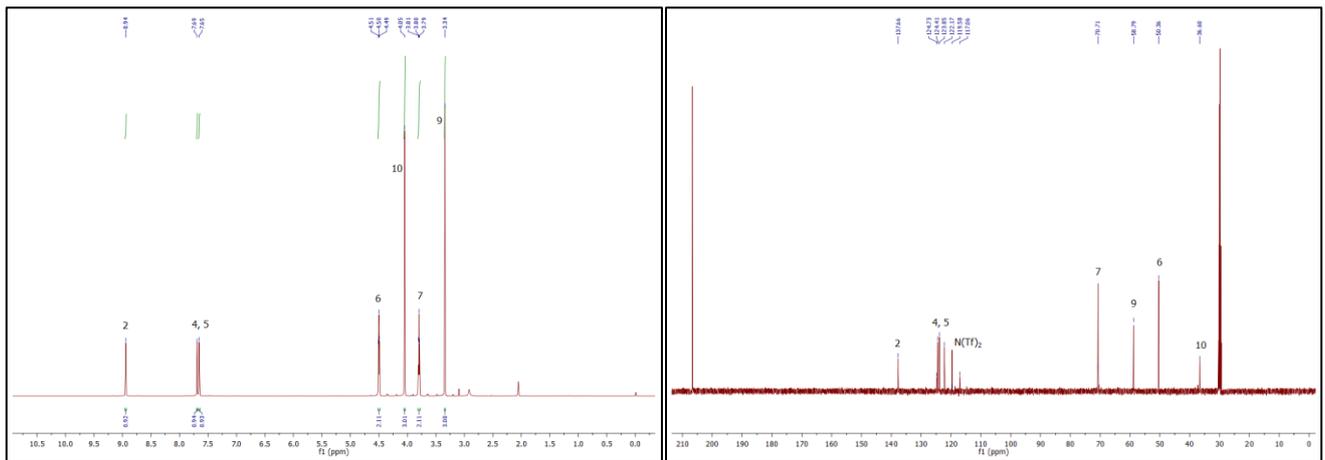

Figure 2: Analyses to IL-NTf a) $^1$H NMR spectrum (500 MHz, 25 °C) in $(CD_3)_2CO$ and b) $^{13}$C NMR spectrum (500 MHz, 25 °C) in $(CD_3)_2CO$.

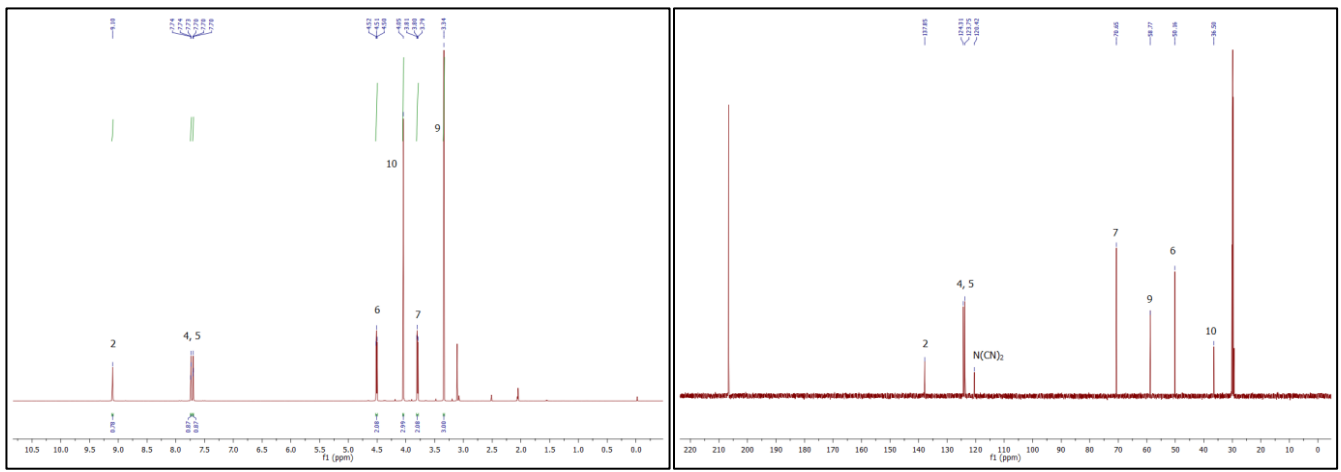

Figure 3: Analyses to IL-NCN a) $^1$H NMR spectrum (400 MHz, 25 °C) in $(CD_3)_2CO$ and b) $^{13}$C NMR spectrum (400 MHz, 25 °C) in $(CD_3)_2CO$.

Nonlocal thermal-optical properties were characterized by the Z-scan technique. The model to a high repetition laser system was present previously by Falconieri [11] and, besides the thermo-optical coefficient dn/dT, it can evaluate properties as the characteristic time of lens formation, $t_c$, and its strength, $\theta$.

Nonlinear optical properties were characterized through the EZ-scan technique since the electronic response is typically small and this technique has a much higher sensibility (Xia, Hagan, and Stryland 1994). As our system consists of a high repetition laser beam, the thermal management technique should be performed [16].

Our experimental setup was calibrated performing a measure of carbon disulfide in a quartz cuvette. The determined nonlinear refractive index was $n_2 = 2.6 \pm 0.3 \times 10^{-15}$ cm$^2$/W, which agrees with the recent data [17].

## 4. Linear and nonlinear characterization

The ILs presented here were characterized by different mechanisms. The linear absorption was evaluated, nonlinear responses and thermo-optical properties were studied.

*1.1 Linear absorption.* The linear absorptions of the ILs were evaluated through spectrum analyses. The measure was performed in a Cary 5000 UV-Vis-NIR from Agilent. The samples were deposited in a $2.0\ mm$ quartz cuvette to execute the analyses. The figure 4 presents the absorption coefficient for the ILs studied in this paper. The narrow inside the graph represents the laser wavelength used in the nonlinear and thermo-optical characterizations. The absorption coefficients, *a*, measured for the wavelength of 780 nm were equals to 0.028 ± 0.010 cm$^{-1}$ for the IL BMIOMe.NTf$_2$ and 0.017 ± 0.015 cm$^{-1}$ for the IL BMIOMe.N(CN)$_2$. The small coefficient values evidence the low absorption of both ILs in the laser wavelength used as the source in the next techniques. These values of the linear absorption in the laser wavelength obtained are also summarized in table 1.

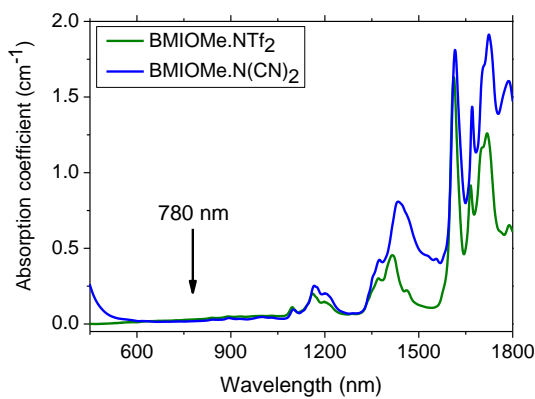

Figure 4: Absorption coefficients for both ionic liquids. The narrow represents the laser wavelength used in the characterization techniques in this work.

*1.2 Nonlinear optical responses.* NLO responses are results of the high orders of polarization excitation. Typically, these orders can be achieved with a peak intensity of the beam comparable with the atomic field, fulfilled condition by some pulsed beam [18].

Nonlinear refractive index, n$_2$, measured in this work, is a fast and local response generated by the excitation of the third order of susceptibility. This property corresponds to the change of the refractive index in the presence of light, i.e., to an intensity-dependent refractive index. In liquids, the nonlinear response results mainly from the deformation generated in the harmonic response of the molecular electronic cloud [18] and by molecular orientation [19].

The nonlinear refractive indexes for ILs were evaluated by the EZ-scan technique. The figures 5a) and c) present the variation of transmittance on sample position generated by the nonlinear refractive index for both ILs. The value obtained to the nonlinear refractive index for the IL BMIOMe.NTf$_2$ was $2.1 \pm 0.4 \times 10^{-16}\ cm^2/W$. No significant signal variation was obtained for the BMIOMe.N(CN)$_2$ sample, which presents a nonlinear refractive index smaller than $3.9 \times 10^{-16}\ cm^2/W$, limited by the sensitivity in that measure. As the BMIOMe.N(CN)$_2$ presents a higher linear absorption coefficient, the laser power used to measure the fast response was reduced, allowing the sample to relax between two sequential temporal windows, even with the small duty cycle. Different anions between the ILs generate distinct optical responses both by electronic and molecular orientation origins. The delocalization of the electron in each molecule and the influence of the anions on the local viscosity implies in a unique response and are responsible for the distinct values of nonlinear refractive indexes. These values are comparable with previous results or with the prevision of others ILs in similar conditions [3,4].

Nonlinear absorption was evaluated by Open Aperture Z-scan technique. This effect occurs when a transition to an excited state from the ground state is achieved by the two-photon absorption process, implying in absorption that has an intensity dependence [18]. A centered valley should be observed in the case of multi-photon absorption since the position z = 0 is the region where the highest intensity is. The results of nonlinear absorption are presented inset of the figures 4a) and b). Evidence of nonlinear absorption was not observed for the ILs in the intensity range here employed for the measurements.

*1.3 Thermo-optical response.* Thermo-optical properties occur by the sample heating promoting a nonlocal response. The thermo-optical coefficient is a result of the variations induced in the density after the linear absorptions and sample heating. Z-scan and EZ-scan setup present a similar position-dependent transmittance profile for both thermal and electronic effects. In many cases, thermal responses are misinterpreted as NLO responses. What strongly differentiate them are the temporal regimes; thermal responses have a slower response and nonlinear are fast. The figures 5b) and d) present the normalized transmittance for both ILs in the EZ-scan technique. These measures were performed at $2.3\ ms$ after the beginning of the beam exposure.

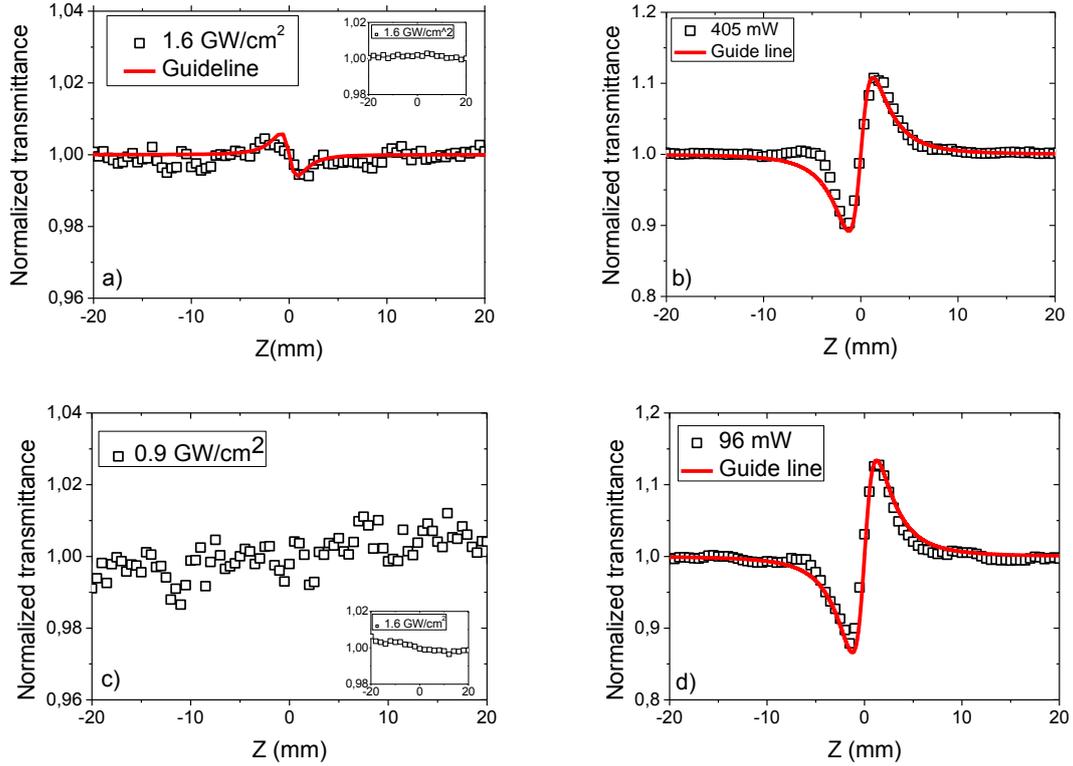

Figure 5: Normalized transmittance of BMIOMe.NTf$_2$ a) at the beginning of exposure temporal window (*t=0*) and b) 2.3 $ms$ of beam exposure, and normalized transmittance of BMIOMe.N(CN)$_2$ c) at the beginning of exposure temporal window (*t=0*) and d) 2.3 $ms$ of beam exposure. The long exposure time characterizes thermal response while $t = 0$ is the electronic response. The correspondent intensities are presented in the graphs. The respectively open aperture Z-scan measurements are inset. Plots b) and d) adapted with permission from ref [20]

Since the model for thermal effects is well-established by Falconieri, the thermo-optical properties were evaluated by the Z-scan technique [11]. The values of the thermo-optical coefficient, dn/dT, thermal characteristic time, $t_c$, and lens strength, $\theta$, were determined by the evolution of the signal in the time. The thermal conductivities of ILs are necessary for thermo-optical coefficient evaluation. To estimate thermal conductivity, $\kappa$, the model proposed by Fröba et al. was used [21]. The respective values of thermal conductivity to ILs are presented in table 1.

Figure 6 presents the temporal evolutions in the peak and valley, pre and post focal, for both ILs measured in the Z-scan technique. The power average beam to both measures was equal to 51 mW in the samples. Note that the graphs do not have the same scale since their thermal optical responses are very different. The red line presents the theoretical fit in each measure.

Adopting the referred thermal management model to this technique, $t_c$ and $\theta$ can be evaluated. These terms are related to the dn/dT by the model proposed by Sheldon et al. [22]. The measured and calculated values are summarized in table 1. The measured and calculated values are summarized in table 1. The characteristic time of lens formation in both IL's are quite similar and comparable to the reported in the literature [23]. It is possible to observe a huge difference in the estimated thermo-optical coefficient between both ILs. Even though BMIOMe.NTf$_2$ has a higher absorption coefficient than BMIOMe.N(CN)$_2$, this later has higher lens strength, thermal conductivity and a higher thermo-optical coefficient consequently. ILs usually presents this last term, dn/dT, in the order of $10^{-5}$ K$^{-1}$ [6], but also can change around two orders of magnitude [7,24], the range which agrees with the results determined in this paper.

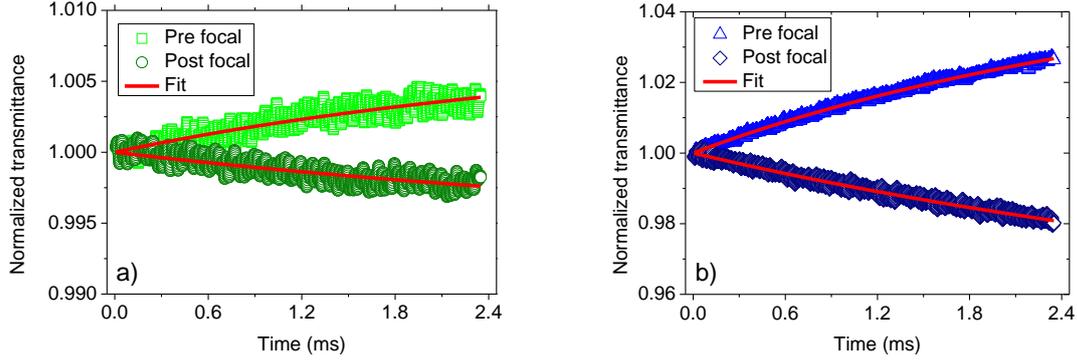

Figure 6: Temporal evolution of normalized transmittance in the position of peak and valley, pre and post focal, of the Z-scan curve for a) the IL BMIOMe.NTf$_2$, tons of green, and b) for the IL BMIOMe.N(CN)$_2$, tons of blue. Both measures were performed with the laser average power in the sample equal to 51 mW.

Table 1: Experimental results for both ILs: Absorption coefficients, $a$, and the nonlinear refractive indexes, $n_2$ at 780nm. Thermal constants; thermal lens strength, θ, thermal characteristic time, $t_c$, thermal conductivity, κ, and the thermo-optical coefficient dn/dT.

| Ionic liquid | $a$ $(cm^{-1})$ | $n_2$ $(10^{-16} cm^2/W)$ | $\theta$ | $t_c$ $(ms)$ | $\kappa$ $(W\ K^{-1}\ m^{-1})$ | $dn/dT\ (K^{-1})$ |
|---|---|---|---|---|---|---|
| BMIOMe.NTf$_2$ | 0.028 ± 0.01 | 2.1 ± 0.4 | 0.021 ± 0.01 | 6.8 ± 1 | 0.111 | $-0.06 \pm 0.01 \times 10^{-6}$ |
| BMIOMe.N(CN)$_2$ | 0.017 ± 0.015 | < 3.9 | 0.16 ± 0.1 | 7.1 ± 1 | 0.197 | $-1.4 \pm 0.7 \times 10^{-5}$ |

## Conclusions

Here we have reported the linear, nonlinear, and thermal optical properties for two ILs; BMIOMe.NTf2 and BMIOMe.N(CN)$_2$, presenting their synthesis procedure. Although extremely significant considering the desired application in photonics, just a few studies have been developed in NLO characterization of ILs. For the first time, the nonlinear refractive index of a pure IL (BMIOMe.NTf2) was determined in the most standard self-refraction setup. This result denotes other possibilities already employed for ILs, where engineered nonlinearities explore additional molecular and plasmonic resonances. The thermo-optical properties for both ILs were characterized; the thermo-optical coefficient, thermal characteristic time, and lens strength. These values obtained agree with the previous researches made in similar and liquids samples.

## Acknowledgments

The authors are greatfull to Conselho Nacional de Desenvolvimento Científico e Tecnológico (CNPq), grant number 142118/2016-8, and Coordenação Aperfeiçoamento de Pessoal de Nível Superior (CAPES), finance code 001.

## References

[1] Bahae M S, Said a a, Wei T H, Hagan D J and Stryland E W Van 1990 Sensitive Measurements of Optical Nonlinearities Using a Single Beam *IEEE J. Quantum Electron.* **26** 760–9

[2] Xia T, Hagan D J, Sheik-Bahae M and Stryland E W Van 1994 Eclipsing Z-scan measurement of A/10 4 wave-front distortion *Opt. Lett.* **19** 317–9

[3] Corrêa N F, Santos C E A, Valadão D R B, de Oliveira L F, Dupont J, Alencar M A R C and Hickmann J M 2016 Third-order nonlinear optical responses of colloidal Ag nanoparticles dispersed in BMIBF_4 ionic liquid *Opt. Mater. Express* **6** 244

[4] Zhao F, Wang C, Zhang J and Zeng Y 2012 Femtosecond third-order optical nonlinearity of an azobenzene-containing ionic liquid crystalline polymer *Opt. Express* **20** 26845

[5] Zhang X, Wang C, Pan X, Xiao S, Zeng Y, He T and Lu X 2010 Nonlinear optical properties and photoinduced anisotropy of an azobenzene ionic liquid – crystalline polymer *Opt. Commun.* **283** 146–50

[6] Souza R F, Alencar M A R C, Meneghetti M R, Dupont J and Hickmann J M 2008 Nonlocal optical nonlinearity of ionic liquids *J. Phys. Condens. Matter* **20** 2–7

[7] Nóvoa-López J A, López Lago E, Domínguez-Pérez M, Troncoso J, Varela L M, De La Fuente R, Cabeza O, Michinel H and Rodríguez J R 2014 Thermal refraction in ionic liquids induced by a train of femtosecond laser pulses *Opt. Laser Technol.* **61** 1–7

[8] Rudenko V, Garbovskiy Y, Klimusheva G and Mirnaya T 2018 Intensity dependent nonlinear absorption coefficients and nonlinear refractive indices of glass-forming ionic liquid crystals doped with gold and silver nanoparticles *J. Mol. Liq.* **267** 56–60

[9] Severiano-Carrillo I, Alvarado-Méndez E, Barrera-Rivera K A, Vázquez M A, Ortiz-Gutierrez M and Trejo-Durán M 2018 Studies of optical nonlinear properties of asymmetric ionic liquids *Opt. Mater. (Amst).* **84** 166–71

[10] Cassol C C, Ebeling G, Ferrera B and Dupont J 2006 A simple and practical method for the preparation and purity determination of halide-free imidazolium ionic liquids *Adv. Synth. Catal.* **348** 243–8

[11] Falconieri M 1999 Thermo-optical effects in Z-scan measurements using high-repetition-rate lasers *J. Opt. A Pure Appl. Opt.* **1** 662–7

[12] Xia T, Hagan D J and Stryland E W Van 1994 Eclipsing Z-scan measurement of lambida/10 4 wave-front distortion *Opt. Lett.* **19** 317–9

[13] Castro H P S, Pereira M K, Ferreira V C, Hickmann J M and Correia R R B 2017 Optical characterization of carbon quantum dots in colloidal suspensions *Opt. Mater. Express* **7** 401

[14] Pereira M K and Correia R R B 2020 Z-scan and eclipsing Z-scan analytical expressions for third-order optical nonlinearities *J. Opt. Soc. Am. B* **37** 478

[15] Ferreira V C, Marin G, Fernandes J A, Dupont J and Correia R R B 2018 Nonlinear optical characterization of new ionic liquids by a noise reduced thermally managed EZ-Scan technique *Nonlinear Opt. its Appl. 2018* 50

[16] Gomes a S L, Filho E L, de Araújo C B, Rativa D and de Araujo R E 2007 Thermally managed eclipse Z-scan. *Opt. Express* **15** 1712–7

[17] Gnoli A, Razzari L and Righini M 2005 Z-scan measurements using high repetition rate lasers: how to manage thermal effects *Opt. Express* **13** 7976

[18] Boyd R W 2003 *Nonlinear Optics* (Weinheim, Germany: Elsevier)

[19] Cang H, Li J and Fayer M D 2003 Orientational dynamics of the ionic organic liquid 1-ethyl-3-methylimidazolium nitrate *J. Chem. Phys.* **119** 13017–23

[20] Ferreira V C, Marin G, Dupont J and Correia R R B 2019 Thermal Analysis of New Ionic Liquids by EZ-Scan Technique *2019 Conference on Lasers and Electro-Optics Europe & European Quantum Electronics Conference (CLEO/Europe-EQEC)* vol Part F140- (IEEE) pp 1–1

[21] Fröba A P, Rausch M H, Krzeminski K, Assenbaum D, Wasserscheid P and Leipertz A 2010 Thermal conductivity of ionic liquids: Measurement and prediction *Int. J. Thermophys.* **31** 2059–77

[22] Sheldon S J, Knight L V and Thorne J M 1982 Laser-induced thermal lens effect: a new theoretical model. *Appl. Opt.* **21** 1663–9

[23] Falconieri M and Salvetti G 1999 Simultaneous measurement of pure-optical and thermo-optical nonlinearities induced by high-repetition-rate, femtosecond laser pulses: Application to CS2 *Appl. Phys. B Lasers Opt.* **69** 133–6

[24] Santos C E A, Alencar M A R C, Migowski P, Dupont J and Hickmann J M 2013 Nonlocal Nonlinear Optical Response of Ionic Liquids under Violet Excitation *Adv. Mater. Sci. Eng.* **2013** 1–6